\documentclass[aps,pra,superscriptaddress,twocolumn,showpacs]{revtex4}

\renewcommand*{\[}{\begin{equation}}

\renewcommand*{\]}{\end{equation}}

\def\PRA{{Phys.~Rev.~A} }

\def\JPB{{J.~Phys.~B} }
\def\PRL{{Phys.~Rev.~Lett.} }

\newcommand{\myscaleboxa}[1]{\scalebox{0.5}[0.5]{#1}}

\newcommand{\myscaleboxc}[1]{\scalebox{1.0}[1.0]{#1}}

\usepackage{epsfig}

\usepackage{hyperref}

\begin{document}

\title{Analysis of effects of macroscopic propagation and multiple molecular orbitals
on the minimum in high-order harmonic generation of aligned CO$_{2}$}

\author{Cheng Jin}

\affiliation{J. R. Macdonald Laboratory, Physics Department, Kansas
State University, Manhattan, Kansas 66506-2604, USA}

\author{Anh-Thu Le}
\affiliation{J. R. Macdonald Laboratory, Physics Department, Kansas
State University, Manhattan, Kansas 66506-2604, USA}

\author{C. D. Lin}
\affiliation{J. R. Macdonald Laboratory, Physics Department, Kansas
State University, Manhattan, Kansas 66506-2604, USA}

\date{\today}

\begin{abstract}
We report theoretical calculations on the effect of the multiple
orbital contribution in high-order harmonic generation (HHG) from
aligned CO$_2$ with inclusion of macroscopic propagation of harmonic
fields in the medium. Our results show very good agreements with
recent experiments for the dynamics of the minimum in HHG spectra as
laser intensity or alignment angle changes. Calculations are carried
out to check how the position of the minimum in HHG spectra depends
on the degrees of molecular alignment, laser focusing conditions,
and the effects of alignment-dependent ionization rates of the
different molecular orbitals. These analyses help to explain why the
minima observed in different experiments may vary.

\end{abstract}


\maketitle

\section{Introduction}
High-order harmonic generation (HHG) is one of the most interesting nonlinear
phenomena when atoms or molecules are exposed to an intense laser field
\cite{Winter-RMP-2008,Krausz-RMP-2000,Agostini-RPP-2004}. Experimentally the
harmonic fields generated by all atoms or molecules within the laser focus
co-propagate with the fundamental laser field till they reach the detector. To
compare with experimental measurements, theoretical harmonics from each atom or
molecule first have to be calculated. The induced dipoles are then fed into the
propagation equation, taking into account the focusing condition, the nature of
the gas medium, and the equation is finally integrated till the harmonics are
collected
\cite{Mette-jpb,Priori-pra-2000,tosa-pra-2005,Geissler-prl-99,jin-pra-2011}.

HHG from randomly distributed molecules have been studied since the
1990's
\cite{Altucci-pra-2006,Wong-pra-2010,Wong-ol-2010,Trallero-CP-2009}.
In recent years, experimental HHG studies tend to focus on partially
aligned molecules
\cite{stanford-Sci-N2,Haessler-NatPhys-2010,Mairesse-prl-2010,Lee-jpb-2010}.
Among the molecules, CO$_2$ is likely the most extensively studied
system so far
\cite{Boutu-NatPhys-2008,Kanai-pra-2008,Wagner-pra-2007,Zhou-prl-2008}.
Initially the interest was focused on the observation of the minimum
in the HHG spectra of CO$_2$
\cite{olga-nature-2009,Vozzi-prl-2005,Torres-pra-2010}. The
positions of the minima from different experiments, however, are
often vastly different. According to the three-step model
\cite{Corkum-prl-1993,Krause-prl-1992}, HHG is generated through the
recombination of the returning electrons with the molecular ion. The
interference of electron waves from the two atomic centers, under
some conditions, results in a minimum in the transition dipole, the
simplest is the two-center interference model
\cite{Kanai-nature-2005,Lein-prl-2002,Le-pra-2006}. In fact,
recombination is an inverse process of photoionization. Thus the
transition dipole is the same as that in photoionization. Any such
minima have not been observed in photoionization, however, since in
these measurements molecules are isotropically distributed.

From the theoretical side, the alignment dependence of HHG was first
studied using the strong-field approximation (SFA) (or the
Lewenstein model)
\cite{Lewen-pra-1994,Zhou-pra-2005a,Zhou-pra-2005b}. Subsequently we
developed a quantitative rescattering (QRS) theory
\cite{lin-jpb-10,toru-2008,at-pra-2009} for HHG which was a
significant improvement on the Lewenstein model. In QRS the accuracy
of recombination is treated at the same level as in the
photoionization process. Using QRS, the HHG minimum is attributed to
the minimum in the photoionization transition dipole between the
highest occupied molecular orbital (HOMO) and continuum molecular
wave functions \cite{at-pra-2009,le-prl-2009}. For fixed-in-space
CO$_2$ molecules the photoionization cross sections of HOMO indeed
exhibit minima at small alignment angles. The experimental HHG
spectra from partially aligned CO$_2$ have been explained reasonably
well by QRS, including the harmonic intensities and the phase, so
are the polarization and ellipticity of the harmonics
\cite{le-prl-2009,at-pra-2010}.

If HHG is generated from HOMO only, then one expects that the
position of the minimum does not significantly change with laser
intensity according to QRS theory. Indeed, strong field ionization
depends exponentially on the ionization potential I$_{\text p}$. The
HOMO-1 and HOMO-2 orbitals in CO$_2$ are 4 and 4.4 eV more deeply
bound than the HOMO \cite{Turner}, thus they are not expected to
contribute significantly to the HHG spectra. However, it is also
well known that tunneling ionization rates depend sensitively on the
symmetry of the molecular orbital \cite{lin-jmo-2006}. The HOMO is a
$\pi_{g}$ orbital. It means that at small alignment angles the
ionization rates are small. For HOMO-2, on the other hand, it is a
$\sigma_{g}$ orbital, thus it has large ionization rate when CO$_2$
molecules are parallel aligned. Thus for small alignment angles,
HOMO-2 may become important even though it is bound 4.4 eV deeper
than the HOMO. (HOMO-1 is a $\pi_{u}$ orbital and thus not expected
to contribute significantly to the HHG.)

The first step of HHG process is tunneling ionization. The alignment
dependence of tunneling ionization rate is usually calculated using
molecular Ammosov-Delone-Krainov (MO-ADK) theory
\cite{tong-pra-2002,zhao-pra-10} or SFA. For most molecules that
have been studied the two models give nearly identical alignment
dependence (after normalization). However, this is not the case for
CO$_2$. Experimentally, the alignment dependence of CO$_2$
ionization reported by Pavi\u{c}i\'{c} {\it et al.}
\cite{Pavi-prl-2007} is very narrowly peaked near alignment angle of
46$^{\circ}$. It differs significantly from the predictions of
MO-ADK and SFA \cite{hoang-jpb-2008}. In fact, so far all
theoretical attempts
\cite{Spanner-pra-2009,Zhao-pra-2009,madsen-pra-2009,chu-pra-2009,Petretti-prl-2010}
have not been able to confirm the sharp alignment dependence
reported in the experiment. Furthermore, the observed HHG spectra
from aligned molecules are inconsistent with the reported
experimental alignment dependence of ionization
\cite{at-pra-2010,at-jpb-2009}.

The HHG spectra of parallel aligned CO$_2$ have been studied in a
number of experiments, with 800-nm lasers \cite{olga-nature-2009} as
well as lasers of longer wavelengths
\cite{Torres-pra-2010,hans-prl-10}. It was observed that the
position of the HHG minimum moved to higher photon energies as the
laser intensity was increased. In Smirnova {\it et al.}
\cite{olga-nature-2009}, such effects were interpreted in terms of
the multi-channel interference (between HOMO and HOMO-2) and the
intricate hole dynamics. Their calculations were based on HHG
generated by a single-molecule response. They introduced filters
into the theory such that only short trajectories contributed to the
signals.  In their calculation, a relative phase between different
channels due to strong-field ionization step was introduced ``by
hand'' in order to obtain the good agreement with the measurement.

As noted at the beginning, to compare theoretically simulated HHG
spectra of molecules with experimental ones, the effect of
macroscopic propagation should be considered. Practically, this has
rarely been done. For atomic targets, propagation effect is usually
included with single-atom induced dipoles calculated using the
Lewenstein model. Only in a few rare occasions the atomic response
is calculated more accurately by solving the time-dependent
Schr\"odinger equation (TDSE) \cite{mette-pra-2006,mette-pra-2002}.
It is also well-known that the Lewenstein model does not predict the
HHG spectra (the intensity) precisely, but the phases of the
harmonics are relatively accurate. This fact is used in the QRS
model, which can be understood as simply replacing the transition
dipole calculated using plane waves in the Lewenstein model by one
calculated using accurate molecular continuum wave functions. The
improvement of single-molecule HHG spectra calculated using QRS has
been well documented in our previous publications
\cite{at-pra-2009,at-jpb-2008}. Computationally, QRS is nearly as
simple as the Lewenstein model (actually QRS is even simpler than
SFA for polyatomic molecules, see Ref. \cite{Zhao-pra-2011}), thus
induced dipole can be easily obtained from QRS to feed into the
propagation equations to account for medium propagation effects.
This has been done for atomic targets for low-laser-intensity and
low-gas-pressure conditions \cite{jin-pra-2009}. It has been
extended recently to the conditions of high intensity and high
pressure for Ar and to molecular targets
\cite{jin-pra-2011,jin-jpb-2011}, including polyatomic molecules
\cite{Zhao-pra-2011}.

In this paper, we report theoretical studies of the propagation
effect on the HHG of CO$_2$, including contributions from HOMO and
HOMO-2. Note that experimentally the effect of propagation on the
HHG spectra has rarely been investigated either, in particular, its
effect on the minimum of the HHG spectra. However, this has changed
recently with the reports of such studies on Ar
\cite{jin-jpb-2011,Farrell-pra-2011,Higuet-2011}. The rest of this
paper is arranged as follows. In Sec. II, we first summarize the
equations used for the propagation calculations. We limit ourselves
to low laser intensity and low pressure. To extend the theory to
high intensity and high pressure, the optical properties of CO$_2$
have to be used and they are not available over a broad range of
frequencies. We then summarize how the single-molecule response from
partially aligned molecules is calculated. In Sec. III the
calculated results are presented. Different factors that can
influence the precise positions of the HHG minima are examined and
reported. These results show that precise theoretical predictions of
the positions of HHG minima in a given experiment is difficult, but
the trend (the direction of the shift of the minimum positions) can
be predicted (or explained). A short summary at the end concludes
this paper.

\section{Theoretical method}
We first briefly describe the theory of propagation of high harmonics in a
macroscopic medium, and in free space, till they reach the detector. To
calculate the induced dipole for individual molecules, we include the
contributions from multiple molecular orbitals. Our method is based on
extending the QRS theory \cite{at-pra-2009}.

\subsection{Propagation of harmonics in the medium}
In general, both the fundamental laser field and the harmonic field are
modified when they co-propagate through a macroscopic medium. In this paper, we
limit ourselves to experiments carried out under the conditions of low laser
intensity and low gas pressure, where the effects of dispersion, Kerr
nonlinearity, and plasma defocusing on the fundamental laser field can be
neglected \cite{jin-pra-2009,jin-pra-2011}. The fundamental laser field is
assumed to be a Gaussian beam in space. Its spatial and temporal dependence can
be expressed in an analytical form \cite{jin-pra-2009}. For high harmonics,
dispersion and absorption effects from the medium are not included. The
dispersion and absorption coefficients depend linearly on gas pressure and
could be ignored under low pressure \cite{jin-pra-2011}. The free-electron
dispersion is also neglected since the plasma frequency is much smaller than
the frequencies of high harmonics \cite{Priori-pra-2000}.

The three-dimensional propagation equation of the harmonic field is
described as
\cite{jin-pra-2011,jin-pra-2009,Priori-pra-2000,Mette-jpb,tosa-pra-2005}

\begin{eqnarray}
\label{harm-freq}\nabla_{\bot}^{2}\tilde{E}_{\text
h}^{\parallel}(r,z',\omega,\alpha)-&&\frac{2i\omega}{c}\frac{\partial
\tilde{E}_{\text h}^{\parallel}(r,z',\omega,\alpha)}{\partial
z'}\nonumber\\&&
=-\omega^{2}\mu_{0}\tilde{P}_{\text{nl}}^{\parallel}(r,z',\omega,\alpha),
\end{eqnarray}
where
\begin{eqnarray}
\tilde{E}_{\text h}^{\parallel}(r,z',\omega,\alpha)=\hat{F}[E_{\text
h}^{\parallel}(r,z',t',\alpha)],
\end{eqnarray}
and
\begin{eqnarray}
\tilde{P}_{\text
{nl}}^{\parallel}(r,z',\omega,\alpha)=\hat{F}[P_{\text
{nl}}^{\parallel}(r,z',t',\alpha)].
\end{eqnarray}
Here $\hat{F}$ is the Fourier transform operator acting on the temporal
coordinate. Eq.~(\ref{harm-freq}) is written in a moving coordinate frame
($z^{\prime}=z$ and $t^{\prime}=t-z/c$) and the term $\partial^{2}E_{\text
h}^{\parallel}/\partial z^{\prime 2}$ is neglected. $\tilde{E}_{\text
h}^{\parallel}(r,z',\omega,\alpha)$ is the parallel component of the harmonic
field with respect to the polarization direction of the probe (or generating)
laser, and $\tilde{P}_{\text {nl}}^{\parallel}(r,z',\omega,\alpha)$ is the
parallel component of the nonlinear polarization caused by the generating laser
field, where $\alpha$ is pump-probe angle, i.e., the angle between the aligning
laser and the harmonic generating laser polarizations.

The nonlinear polarization term can be expressed as
\begin{eqnarray}
\label{pola}\tilde{P}_{\text{nl}}^{\parallel}(r,z',\omega,\alpha)=\hat{F}{\{[n_{0}-n_{\text
e}(r,z',t',\alpha)]D^{\parallel,\text{tot}}(r,z',t',\alpha)\}},
\nonumber\\
\end{eqnarray}
where $n_{0}$ is the density of neutral molecules,
$D^{\parallel,\text{tot}}(t',\alpha)$ is the parallel component of
the induced single-molecule dipole over a number of active electrons
[see Eq.~(\ref{total-dip}) below], and $n_{\text e}(t',\alpha)$ is
the free-electron density, which can be calculated as following:
\begin{eqnarray}
\label{free-afa}n_{\text e}(t',\alpha)=\int^{\pi}_{0}n_{\text
e}(t',\theta)\rho(\theta,\alpha)\sin\theta d\theta.
\end{eqnarray}
Here $n_{\text e}(t',\theta)$ is the alignment-dependent
free-electron density, obtained from
\begin{eqnarray}
\label{free-electron}n_{\text
e}(t',\theta)=n_{0}\{1-\exp[-\int_{-\infty}^{t'}\gamma(\tau,\theta)d\tau]\},
\end{eqnarray}
where $\gamma(\tau,\theta)$ is the alignment-dependent ionization
rate, which can be calculated by MO-ADK theory
\cite{tong-pra-2002,zhao-pra-10} for different molecular orbitals.
In Eq.~(\ref{free-afa}), $\theta$ is the alignment angle with
respect to the polarization direction of the probe laser, and
$\rho(\theta,\alpha)$ is the alignment distribution with pump-probe
angle $\alpha$ \cite{jin-pra-2010,at-pra-2009,lein-jmo-2005}.

After the propagation in the medium, we obtain the parallel
component of near-field harmonics on the exit face of the gas jet
($z'=z_{\text {out}}$). For isotropically distributed molecules and
partially aligned molecules with $\alpha=0^{\circ}$ or $90^{\circ}$,
by symmetry there are only parallel harmonic components. The
perpendicular components, which are usually much smaller, would
appear for partially aligned molecules and the harmonics will be
elliptically polarized in general \cite{at-pra-2010}. Generalization
of Eq.~(\ref{harm-freq}) to the perpendicular component is
straightforward, but we restrict ourselves to parallel component
only. Eq.~(\ref{harm-freq}) is solved numerically using a
Crank-Nicholson routine for each value of $\omega$. Typical
parameters used in the calculations are 300 grid points along the
radial direction and 400 grid points along the longitudinal
direction.

\subsection{Harmonics in the far field}
Once the near-field harmonics are obtained on the exit face of the
medium, they further propagate in free space before detected by the
spectrometer. In the meanwhile, they may pass through a slit, an
iris, or be reflected by a mirror. The far-field harmonics can be
obtained from near-field harmonics through a Hankel transformation
\cite{far-field,L'Huillier-1992,tosa-2009}

\begin{eqnarray}
\label{far-hhg}{E}_{\text h}^{\text f}(r_{\text f},z_{\text
f},\omega,\alpha)=&&-ik\int\frac{\tilde{E}_{\text
h}^{\parallel}(r,z',\omega,\alpha)}{z_{\text f}-z'}J_{0}(\frac{k
rr_{\text f}}{z_{\text f}-z'})\nonumber\\&&\times \exp[\frac{i
k(r^{2}+r_{\text f}^{2})}{2(z_{\text f}-z')}] r dr,
\end{eqnarray}
where $J_{0}$ is the zero-order Bessel function, $z_{\text f}$ and $r_{\text f}$
are the coordinates of the far-field points. The wave vector $k$
is given by $k=\omega/c$.

We assume that the harmonics in the far field are collected from an
extended area. By integrating harmonic yields over the area, the
power spectrum of the macroscopic harmonics is obtained by
\begin{eqnarray}
\label{total-hhg}S_{\text
h}(\omega,\alpha)\propto\int\int|{E}_{\text h}^{\text f}(x_{\text
f},y_{\text f},z_{\text f},\omega,\alpha)|^{2}dx_{\text f}dy_{\text
f},
\end{eqnarray}
where $x_{\text f}$ and $y_{\text f}$ are the Cartesian coordinates
on the plane perpendicular to the propagation direction, and
$r_{\text f}=\sqrt{x_{\text f}^{2}+y_{\text f}^{2}}$. To simulate
experimental HHG spectra quantitatively, besides laser parameters,
detailed information on the experimental setup is needed.

\subsection{Quantitative rescattering theory for a multi-orbital molecular system}
In Eq.~(\ref{pola}), laser induced single-molecule dipole moment
$D(t')$ is calculated quantum mechanically using the QRS theory,
which has been discussed in detail for molecules in Ref.
\cite{at-pra-2009}. Within QRS, laser induced dipole moment
$D(\omega,\theta)$ for a molecule at a fixed angle $\theta$
(measured with respect to laser polarization) is given by
\begin{eqnarray}
\label{mol-qrs}D^{\parallel,\perp}(\omega,\theta)=N(\theta)^{1/2}
W(\omega)d^{\parallel,\perp}(\omega,\theta),
\end{eqnarray}
where $N(\theta)$ is the alignment-dependent ionization probability,
$W(\omega)$ is the recombining electron wave packet, and
$d^{\parallel,\perp}(\omega,\theta)$ is the parallel component (or
perpendicular component) of the photorecombination (PR) transition dipole
(complex in general). Here we only consider   linearly polarized lasers and
linear molecules. $W(\omega)$ is independent of the alignment angle $\theta$.
From Eq. (9), it can be expressed as
\begin{eqnarray}
\label{mol-wave}W(\omega)=\frac{D^{\parallel,\perp}(\omega,\theta)}
{N(\theta)^{1/2}d^{\parallel,\perp}(\omega,\theta)}.
\end{eqnarray}

In QRS, $W(\omega)$ is usually calculated only once for a given
angle $\theta$ using SFA, where all the elements on the right-hand
side of Eq.~(\ref{mol-wave}) are obtained by replacing the continuum
waves with plane waves. In QRS, accurate
$d^{\parallel,\perp}(\omega,\theta)$ are obtained from quantum
chemistry code \cite{Lucchese-co2-82,Lucchese-n2-82} and tunneling
ionization probability $N(\theta)$ are obtained either from MO-ADK
\cite{tong-pra-2002,zhao-pra-10} or from SFA. Put all of these
together into Eq.~(\ref{mol-qrs}), laser induced dipole moment
$D(\omega,\theta)$ for each orbital is obtained. Note that in QRS,
the wave packet $W(\omega)$ automatically includes the phase which
is dependent on the molecular orbital. Thus there is no need to
introduce the relative phase between orbitals, in contrast to the
approach in Ref. \cite{olga-nature-2009}. SFA is used to calculate
the ionization probability in Eq.~(\ref{mol-qrs}) in this paper
unless otherwise stated.

We assume that the degree of molecular alignment is not varied
spatially within the medium because the molecules are usually
partially aligned by a weak and loosely focused laser beam
\cite{jin-pra-2010}. By coherently averaging the induced dipole
moments over the molecular angular distribution, we obtain the
averaged induced dipole of partially aligned molecules at each
point inside the medium:
\begin{eqnarray}
\label{avg-in-dip}D^{\parallel,\text
{avg}}(\omega,\alpha)=\int^{\pi}_{0}D^{\parallel}(\omega,\theta)\rho(\theta,\alpha)\sin\theta
d\theta,
\end{eqnarray}
where $\rho(\theta,\alpha)$ is again the angular (or alignment)
distribution of the molecules with respect to the polarization
direction of the probe laser. For randomly distributed molecules,
$\rho(\theta,\alpha)$ is a constant. Note that the above procedure
is only for the specified molecular orbital.

For the specific problem addressed in this paper, we consider electrons either
in the HOMO (1$\pi_{g}$) or in the HOMO-2 (3$\sigma_{u}$) of CO$_{2}$. The
electrons are freed and then recombine back to the same orbital. As discussed
above, the multiple orbital effects are important at small alignment angles
only due to symmetry consideration. At these angles HOMO-1 (1$\pi_{u}$) is
ignored since the ionization rate of HOMO-1 is quite small compared with HOMO
and HOMO-2 \cite{zhao-pra-10,Spanner-pra-2009,olga-nature-2009}. At large
alignment angles, only HOMO becomes important. The total laser induced dipole
over a number of active electrons can be written as
\cite{Madsen-2006,Faria-pra-2010}
\begin{eqnarray}
\label{total-dip}D^{\parallel,\text
{tot}}(\omega,\alpha)=\sum_{j,n}D_{j,n}^{\parallel,\text
{avg}}(\omega,\alpha),
\end{eqnarray}
where index $j$ refers to the HOMO and HOMO-2 of CO$_{2}$, and $n$ is an index
to account for degeneracy in each molecular orbital.

Before proceeding, we comment that the QRS theory is formulated in
the energy (or frequency) domain. There is no explicit treatment of
``core dynamics" as in Ref. \cite{olga-nature-2009}. The time
evolution of the core is reflected only by the energy of the core in
the free propagation of the electron wave packet. In other words,
any possible many-body interchannel couplings between HOMO and
HOMO-2 are not included in the present treatment. Before such
effects are addressed, other factors that are more important on the
HHG as discussed here have to be treated first.

For the propagation of harmonics in the medium, we need to obtain
hundreds of the total induced dipoles $D^{\parallel,\text
{tot}}(\omega,\alpha)$ for different laser intensities, and then
they are fed into Eq.~(\ref{harm-freq}). The same procedure is used
in Jin {\it et al.} \cite{jin-pra-2009} for atomic targets. Note
that in the present model, dielectric properties of molecules due to
non-isotropic distributions are also neglected.

\section{Results and discussion}

\subsection{HHG spectra of randomly distributed CO$_{2}$: theory vs experiment}
\begin{figure*}
\mbox{\rotatebox{270}{\myscaleboxa{
\includegraphics{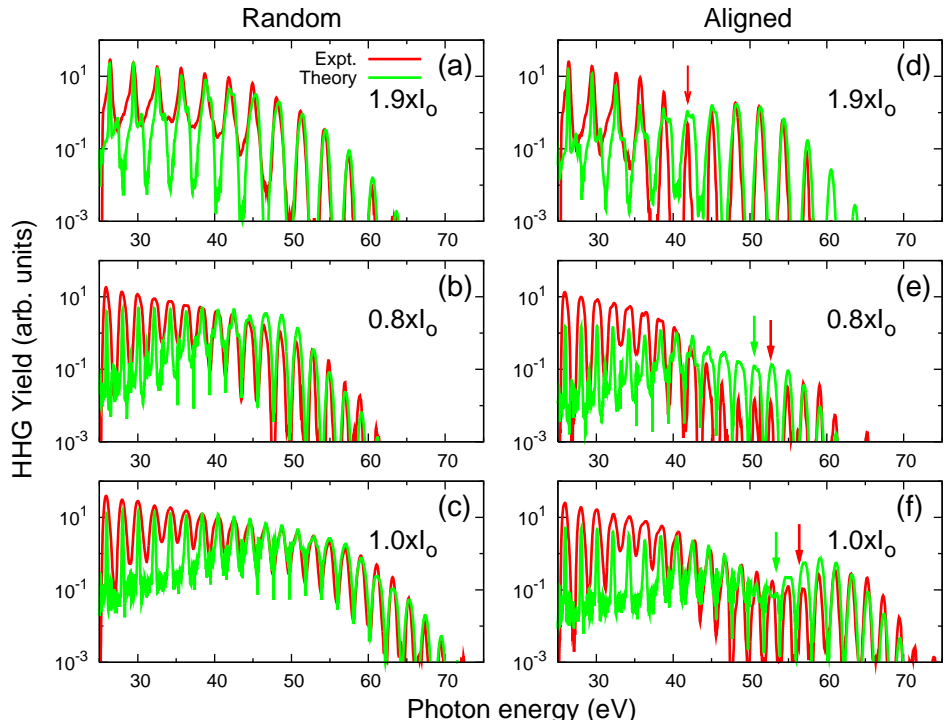}}}}
\caption{(Color online) Comparison of theoretical (green lines) and
experimental (red lines) HHG spectra of random and aligned CO$_{2}$
molecules, in an 800-nm laser shown in (a) and (d), and in a 1200-nm
laser shown in (b), (c), (e), and (f). Laser intensities are
indicated where I$_{\circ}$=10$^{14}$ W/cm$^{2}$. Experimental data
are from Ref. \cite{hans-prl-10}. Arrows indicate the positions of
minima. Pump-probe angle $\alpha$=0$^{\circ}$. See text for
additional laser parameters and experimental arrangements.
\label{Fig1}}
\end{figure*}

HHG spectra by 800-nm and 1200-nm lasers have been reported for
isotropically distributed and partially aligned N$_{2}$ and CO$_{2}$
molecules \cite{hans-prl-10}. The spectra of N$_{2}$ have been
analyzed by Jin {\it et al.} \cite{jin-pra-2011,jin-jpb-2011}
recently including only the HOMO.

In Figs.~\ref{Fig1}(a)-\ref{Fig1}(c), we show the HHG spectra for
isotropically distributed CO$_{2}$ molecules by 800-nm and 1200-nm
lasers. To obtain good agreement between theory and experiment,
especially in the cutoff region, the intensity used in the theory is
adjusted from the value given in the experiment. In
Figs.~\ref{Fig1}(a)-\ref{Fig1}(c), the intensities in theory
(experiment) are 1.9 (2.1), 0.8 (1.0), 1.0 (1.2), in units of
10$^{14}$W/cm$^{2}$, respectively. Other parameters used in the
simulation are the same as those given in the experiment
\cite{hans-prl-10}. The laser parameters are: pulse duration is
$\sim$ 32 fs (800 nm) or $\sim$ 44 fs (1200 nm), beam waist at the
focus is $\sim$ 40 $\mu$m. A 0.6-mm-wide gas jet is located 3 mm
(800 nm) or 3.5 mm (1200 nm) after the laser focus, and a slit with
a width of 100 $\mu$m is placed at 24 cm after the gas jet.

Figs.~\ref{Fig1}(a)-\ref{Fig1}(c) clearly show the good overall
agreement between experiment and theory for randomly distributed
CO$_{2}$ molecules. We have checked that HOMO is dominant for
randomly distributed CO$_2$, with negligible contributions from
inner orbitals. The HHG spectra do not exhibit any minima, as
opposed to the spectra of random N$_{2}$ molecules when they are
generated under the same experimental conditions \cite{hans-prl-10}.
For randomly distributed CO$_{2}$, there was no minimum found in HHG
spectra using an 800-nm laser by Vozzi {\it et al.}
\cite{Vozzi-prl-2005}. However, for the 1300-nm lasers, the data
from Torres {\it et al.} \cite{Torres-pra-2010,Torres-oe-2010}
appear to show a very weak minimum at photon energy near 45 eV.
Without more careful study including different intensities and
wavelengths, however, this is not conclusive.

\subsection{ HHG spectra of aligned CO$_{2}$: theory vs experiment}

Experimentally HHG spectra have also be reported from aligned
CO$_{2}$ molecules. A relatively weak and short laser pulse was used
to impulsively align molecules, and the HHG spectra were taken at
half-revival ($\sim$ 21.2 ps in CO$_{2}$) when the molecules were
maximally aligned \cite{hans-prl-10}. The angular distributions of
the aligned molecules are obtained by solving the TDSE of rotational
wave packet \cite{jin-pra-2010}. The degree of alignment is
$\langle\cos^{2}\theta\rangle$=0.60 in Fig.~\ref{Fig1}(d), and
$\langle\cos^{2}\theta\rangle$=0.50 in Figs.~\ref{Fig1}(e)
and~\ref{Fig1}(f). The polarizations of the pump and probe lasers
are parallel to each other.

The HHG spectra of partially aligned CO$_{2}$ molecules are shown in
Figs.~\ref{Fig1}(d)-\ref{Fig1}(f), which are obtained under the same
generating lasers and experimental arrangements as those in
Figs.~\ref{Fig1}(a)-\ref{Fig1}(c), respectively. The simulation and
experimental data agree well with each other in general. In
Fig.~\ref{Fig1}(e), the discrepancy is a little bigger, showing the
drop near 40 eV is larger in the experiment than in the theory. But
we note that in Fig.~\ref{Fig1}(f), the experimental data do not
drop as rapidly, in agreement with the theoretical simulation.

The minima in the HHG spectra of CO$_2$ and their dependence on laser intensity
have been widely discussed in the literature
\cite{Torres-pra-2010,olga-nature-2009}. In Fig.~\ref{Fig1}(d), for an 800-nm
laser, experiment gives a strong minimum at 42$\pm$2 eV, our simulation
predicts a minimum around 42 eV. For the 1200-nm laser, in Fig.~\ref{Fig1}(e),
experiment shows a minimum at 51$\pm$2 eV, theory predicts a minimum around 50
eV. In Fig.~\ref{Fig1}(f), the experimental minimum is shifted to 57$\pm$2 eV,
and the theoretical one is moved to around 53.5 eV. Thus our simulation also
shows the shift of the minimum from low to high harmonic orders as laser
intensity is increased. Below we interpret the origin of the  shift.

\subsection{Origin of minimum in the HHG spectra of aligned CO$_{2}$: Type I and Type II}
\begin{figure*}
\mbox{\rotatebox{270}{\myscaleboxa{
\includegraphics{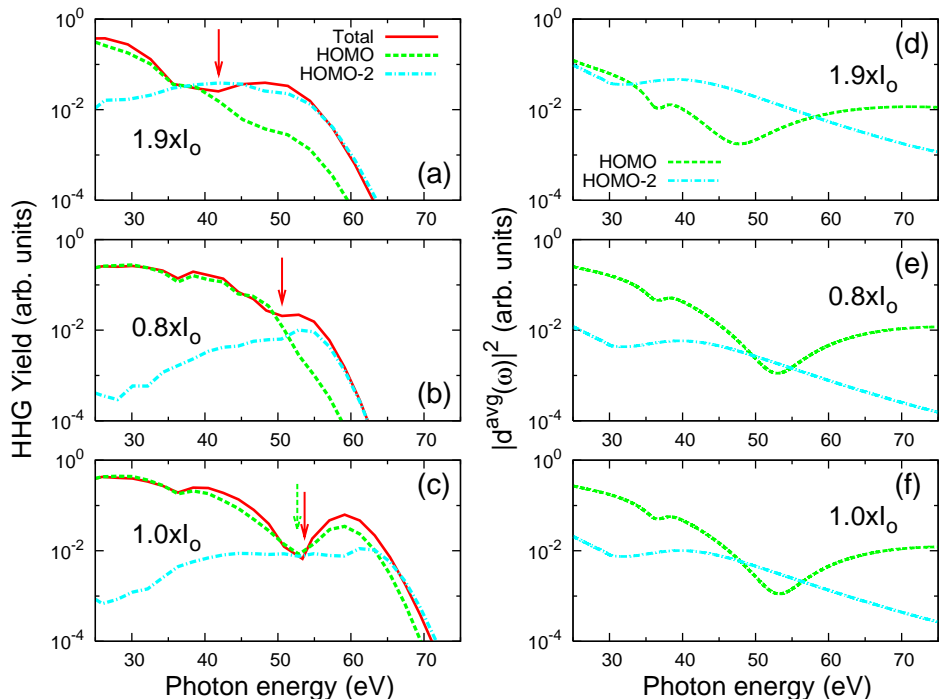}}}}
\caption{(Color online) (a)-(c) Macroscopic HHG spectra (envelope
only) corresponding to Figs.~\ref{Fig1}(d)-\ref{Fig1}(f),
respectively. Total (HOMO and HOMO-2 together) spectra and the
spectra of individual HOMO and HOMO-2 are shown. (d)-(f) Averaged
photorecombination transition dipoles (parallel component, the
square of magnitude) of HOMO and HOMO-2 corresponding to (a), (b),
and (c), respectively. Laser intensities are indicated where
I$_{\circ}$=10$^{14}$ W/cm$^{2}$. Arrows indicate the positions of
minima. Pump-probe angle $\alpha$=0$^{\circ}$. \label{Fig2}}
\end{figure*}

We next analyze the origin of the minimum in the HHG spectra seen in
Figs.~\ref{Fig1}(d)-\ref{Fig1}(f), and consider the dominant contributions from
HOMO and HOMO-2 only. First we define the averaged PR transition dipole for
each molecular orbital by
\begin{eqnarray}
\label{dip-avg}d^{\parallel,\text
{avg}}(\omega,\alpha)=\int^{\pi}_{0}N(\theta)^{1/2}
d^{\parallel}(\omega,\theta)\rho(\theta,\alpha)\sin\theta
d\theta.
\end{eqnarray}

This gives a measure of the relative contribution of each molecular
orbital to the HHG, which is obtained by averaging over the angular
(or alignment) distribution of the partially aligned molecules,
weighted by the square root of the tunneling ionization probability.
The relative ionization rates between HOMO and HOMO-2 change with
laser intensity.

Figs.~\ref{Fig2}(a)-\ref{Fig2}(c) show the envelopes of the HHG
spectra from individual molecular orbitals together with the total
ones, each obtained after propagation in the medium. In the
meanwhile, the averaged PR transition dipoles of HOMO and HOMO-2
under different generating lasers and alignment distributions are
shown in Figs.~\ref{Fig2}(d)-\ref{Fig2}(f), respectively.

In Figs.~\ref{Fig2}(a) and~\ref{Fig2}(b), there are no minima in the
HHG spectra of HOMO or HOMO-2, but the minimum shows up in the total
spectra. This is due to the interference between HOMO and HOMO-2. We
call this type I minimum. Clearly the minimum position will change
with laser intensity since the relative ionization rates between
HOMO and HOMO-2 change with intensity [also see
Figs.~{\ref{Fig5}}(c) and {\ref{Fig5}}(d)]. Similar analysis can be
found in Ref. \cite{olga-nature-2009}. In Fig.~\ref{Fig2}(c), there
is a minimum in the HOMO spectra at 52.6 eV. This minimum is shifted
to 53.6 eV in the total spectra due to the interference with the
HOMO-2. This is categorized as type II minimum. Similar analysis of
this type can be found in Refs. \cite{Torres-pra-2010,olga-pnas-09}.
The minimum in the HOMO spectra is due to the minimum in the
averaged PR transition dipole of HOMO shown in Fig.~\ref{Fig2}(f).
But their positions differ due to modification of the macroscopic
wave packet (MWP). In this connection we mention that the earlier
calculations \cite{le-prl-2009,at-pra-2009} with an 800-nm laser
showed minimum in HHG spectra at small pump-probe angles due to the
contribution from the HOMO only. These calculations were carried out
with a higher degree of alignment and higher laser intensities as
compared to our present study. This is expected as the minimum in
the averaged PR transition dipole from HOMO in Fig.~\ref{Fig2}(d)
becomes deeper and is slightly shifted away from the cutoff to lower
energies with increased degree of alignment [see Fig.~\ref{Fig2}(e)
and \ref{Fig2}(f)]. Furthermore, as shown in Ref.
\cite{jin-jpb-2011}, the minimum in the HOMO spectra could also
result from the multiplication of MWP and averaged PR transition
dipole even when neither has minimum. When a minimum occurs in the
dominant orbital, its position will not change much if it remains
the dominant one when the laser intensity changes. The little bump
around 36 eV in the HOMO spectra as well as in the total spectra can
be seen due to the bump in the HOMO curves in
Figs.~\ref{Fig2}(d)-\ref{Fig2}(f). Its position does not change much
since the HOMO-2 remains small.

Previously in Refs. \cite{jin-jpb-2011,jin-pra-2011}, we have shown
that the macroscopic HHG spectrum is the product of a MWP and an
averaged PR transition dipole for each individual molecular orbital.
Since the ionization rate for each orbital has been incorporated in
the averaged PR transition dipole, the MWP is mostly identical
except for the phase due to ionization potential. The averaged PR
transition dipole is very sensitive to ionization rates. The
relative magnitude changes rapidly with the increase of laser
intensity. Thus when two averaged PR transition dipoles are
comparable [see Fig.~\ref{Fig2}(d)], the position of the minimum
changes rapidly with the laser intensity. The averaged PR transition
dipole is also sensitive to alignment distributions [see
Figs.~\ref{Fig2}(d)-\ref{Fig2}(f)]. At low laser intensity, HOMO-2
is small, interference often occurs in a narrow region only where
the two amplitudes are comparable, see Figs.~\ref{Fig2}(b) and
\ref{Fig2}(c). In comparison, in Smirnova {\it et al.}
\cite{olga-nature-2009}, HOMO and HOMO-2 tend to interfere over a
broad photon-energy region. The ionization rates and transition
dipoles used in their calculations are different from ours.

\subsection{Progression of harmonic minimum vs laser intensity}
\begin{figure}
\mbox{\rotatebox{270}{\myscaleboxa{
\includegraphics{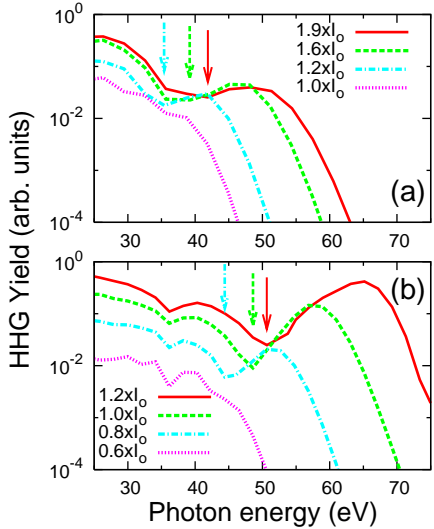}}}}
\caption{(Color online) Laser intensity dependence of macroscopic HHG spectra
(envelope only)   (a) in an 800-nm laser and (b) in a 1200-nm laser.
Intensities are shown in units of I$_{\circ}$=10$^{14}$ W/cm$^{2}$. Arrows
indicate the positions of minima. Degree of alignment:
$\langle\cos^{2}\theta\rangle$=0.60. Pump-probe angle $\alpha$=0$^{\circ}$.
\label{Fig3}}
\end{figure}

In Figs.~{\ref{Fig3}}(a) and {\ref{Fig3}}(b) we show the envelope of
the calculated HHG spectra for four different peak intensities with
an 800-nm laser and a 1200-nm laser, respectively. For the 800-nm
spectra, the lowest intensity does not have a minimum. For the
higher ones, each spectrum has a type I minimum, with its position
shifts to lower photon energy as the laser intensity is decreased.
The degree of alignment of molecules used in the calculation is
$\langle\cos^{2}\theta\rangle$=0.60. We find that the shift cannot
be attributed to either MWP or the averaged PR transition dipole
alone. For the 1200-nm data, also with
$\langle\cos^{2}\theta\rangle$=0.60, which is different from
Figs.~{\ref{Fig1}}(e) and {\ref{Fig1}}(f), we find that the minimum
is type II, where the averaged PR transition dipole of the HOMO has
a minimum. The minimum in the HHG spectra of the HOMO shifts to
higher photon energy as the intensity increases, but the
interference with HOMO-2 shifts the minimum to even higher energies.
In other words, the shift of the position of the HHG minimum vs
intensity cannot be attributed to a single factor alone.

\subsection{Other factors that influence the precise positions of HHG minima}
\begin{figure*}
\mbox{\rotatebox{270}{\myscaleboxa{
\includegraphics{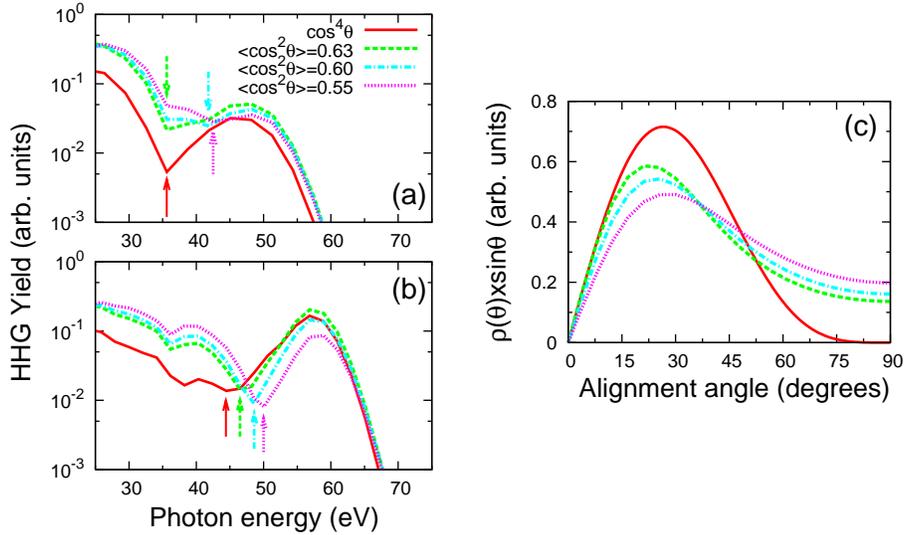}}}}
\caption{(Color online) Dependence of macroscopic HHG spectra
(envelope only) with degrees of molecular alignment distributions
for (a) an 800-nm laser with intensity of 1.8$\times$10$^{14}$
W/cm$^{2}$, and (b) a 1200-nm laser with intensity of
1.0$\times$10$^{14}$ W/cm$^{2}$. The weighted angular distributions
of the molecules are shown in (c). Arrows indicate the positions of
minima. Pump-probe angle $\alpha$=0$^{\circ}$.\label{Fig4}}
\end{figure*}

In our analysis, the averaged PR transition dipole is calculated over the
angular distribution of the molecules and thus depend on the degrees of
alignment. Since the latter cannot be accurately measured, we check how
sensitive the calculated spectra is with respect to the assumed alignment
distribution. In Fig.~{\ref{Fig4}}(c), four different alignment distributions
are shown. The distributions are multiplied by the volume element $\sin\theta$
for easy comparison. Three of them are obtained from the calculated rotational
wave packets \cite{jin-pra-2010}, with $\langle\cos^{2}\theta\rangle$ as 0.63,
0.60, and 0.55, respectively. The other is the commonly used $\cos^{4}\theta$
distribution. For 800-nm and 1200-nm lasers, the envelopes of the calculated
HHG spectra are shown in Figs.~{\ref{Fig4}}(a) and {\ref{Fig4}}(b),
respectively. The precise position of the minimum changes slightly except for
the one from the $\cos^{4}\theta$ distributions. However, change of a couple of
eV's is seen.

\begin{figure*}
\mbox{\rotatebox{0}{\myscaleboxc{
\includegraphics{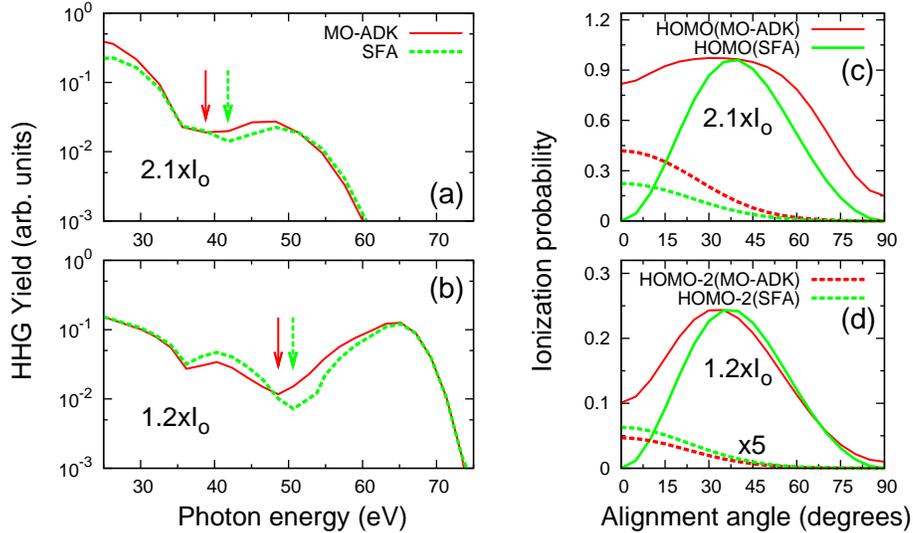}}}}
\caption{(Color online) Dependence of macroscopic HHG spectra
(envelope only) on the ionization probabilities calculated from
MO-ADK or SFA in (a) an 800-nm laser, and (b) a 1200-nm laser. Laser
intensities are indicated where I$_{\circ}$=10$^{14}$ W/cm$^{2}$.
Arrows indicate the positions of minima. Degree of alignment:
$\langle\cos^{2}\theta\rangle$=0.60. Pump-probe angle
$\alpha$=0$^{\circ}$. (c) and (d) Alignment-dependent ionization
probabilities of HOMO and HOMO-2 calculated using MO-ADK and SFA.
Laser parameters are the same as (a) and (b). Ionization
probabilities of HOMO-2 in (d) are multiplied by 5. \label{Fig5}}
\end{figure*}

To precisely determine the minimum in the HHG spectra, accurate
alignment-dependent ionization probability $N(\theta)$ for each
molecular orbital is needed. For CO$_{2}$, even for HOMO, different
theories in the literature
\cite{zhao-pra-10,Spanner-pra-2009,olga-nature-2009,Petretti-prl-2010,
Pavi-prl-2007,hoang-jpb-2008,madsen-pra-2009,chu-pra-2009} show
non-negligible differences, and they do not agree with the
experimental data \cite{Pavi-prl-2007}. Here we examine how the HHG
spectra change with the different ionization rates used. The
ionization rates for both HOMO and HOMO-2 can all be easily
calculated from SFA or from MO-ADK theory. Figs.~{\ref{Fig5}}(a) and
{\ref{Fig5}}(b) show the HHG spectra calculated using the ionization
probabilities shown in Figs.~{\ref{Fig5}}(c) and {\ref{Fig5}}(d).
Other laser parameters used in the calculation are given in the
figure captions. The difference of the position of the minimum is 3
eV in Fig.~{\ref{Fig5}}(a) and 2 eV in Fig.~{\ref{Fig5}}(b). Note
that the ionization probabilities from SFA and MO-ADK are normalized
at the peak of the HOMO curve. In Fig.~{\ref{Fig5}}(a) the spectra
are normalized at H33 (51 eV) and in Fig.~{\ref{Fig5}}(b) at H65 (67
eV).

\begin{figure}
\mbox{\rotatebox{270}{\myscaleboxa{
\includegraphics{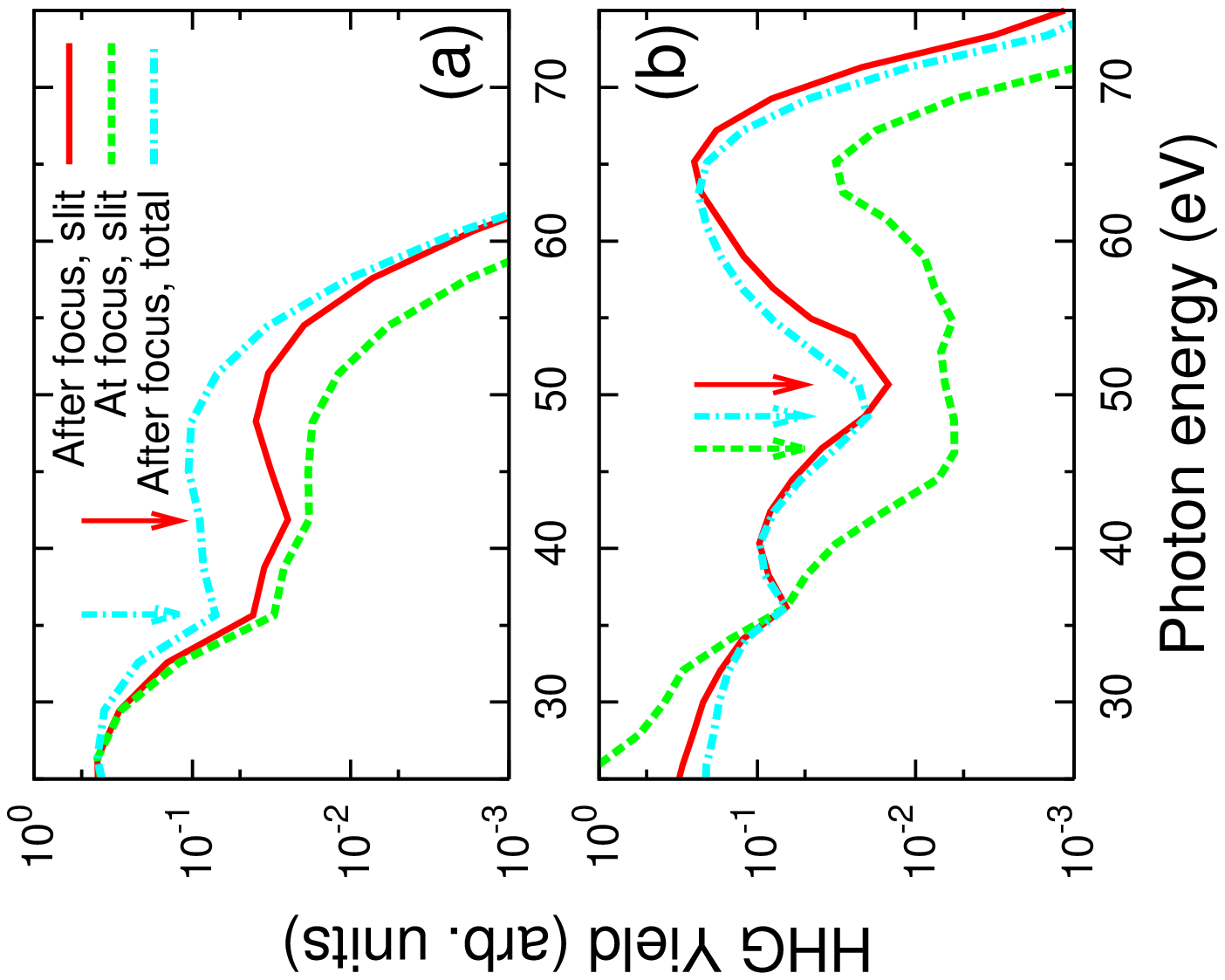}}}}
\caption{(Color online) Dependence of macroscopic HHG spectra
(envelope only) on experimental arrangements (a) for an 800-nm laser
with intensity of 2.1$\times$10$^{14}$ W/cm$^{2}$, and (b) for a
1200-nm laser with intensity of 1.2$\times$10$^{14}$ W/cm$^{2}$. The
arrangements are: (1) gas jet after focus and slit is used (solid
lines); (2) gas jet at the focus and slit is used (dashed lines);
and (3) gas jet is after the focus but without the slit (dot-dashed
lines). Arrows indicate the positions of minima. Degree of
alignment: $\langle\cos^{2}\theta\rangle$=0.60. Pump-probe angle
$\alpha$=0$^{\circ}$. \label{Fig6}}
\end{figure}

The HHG spectra are also sensitive to the experimental arrangement
and thus can also move the position of the HHG minimum. To
demonstrate this, we (i) move the gas jet to the laser focus and
collect the signal using a slit; (ii) put the gas jet after the
laser focus, and collect HHG signal without slit (total signal).
These two will be compared to the arrangement used in this paper:
gas jet is after the laser focus and the HHG is collected with a
slit. The results are compared in Fig.~{\ref{Fig6}}. Note that the
spectra are normalized at H17 (26 eV) in Fig.~{\ref{Fig6}}(a) and at
H35 (36 eV) in Fig.~{\ref{Fig6}}(b). Not only the spectra change
quite significantly, but also the position of the HHG minimum. This
illustrates that it is very difficult to compare the position of the
HHG minimum from different experiments. In this comparison, the
change of the HHG spectra is due to the change of MWP which depends
on the experimental setup. The averaged PR transition dipoles of
HOMO and HOMO-2 are the same in the three calculations.

\subsection{The dependence of the HHG minimum on the pump-probe angle}
\begin{figure*}
\mbox{\rotatebox{270}{\myscaleboxa{
\includegraphics{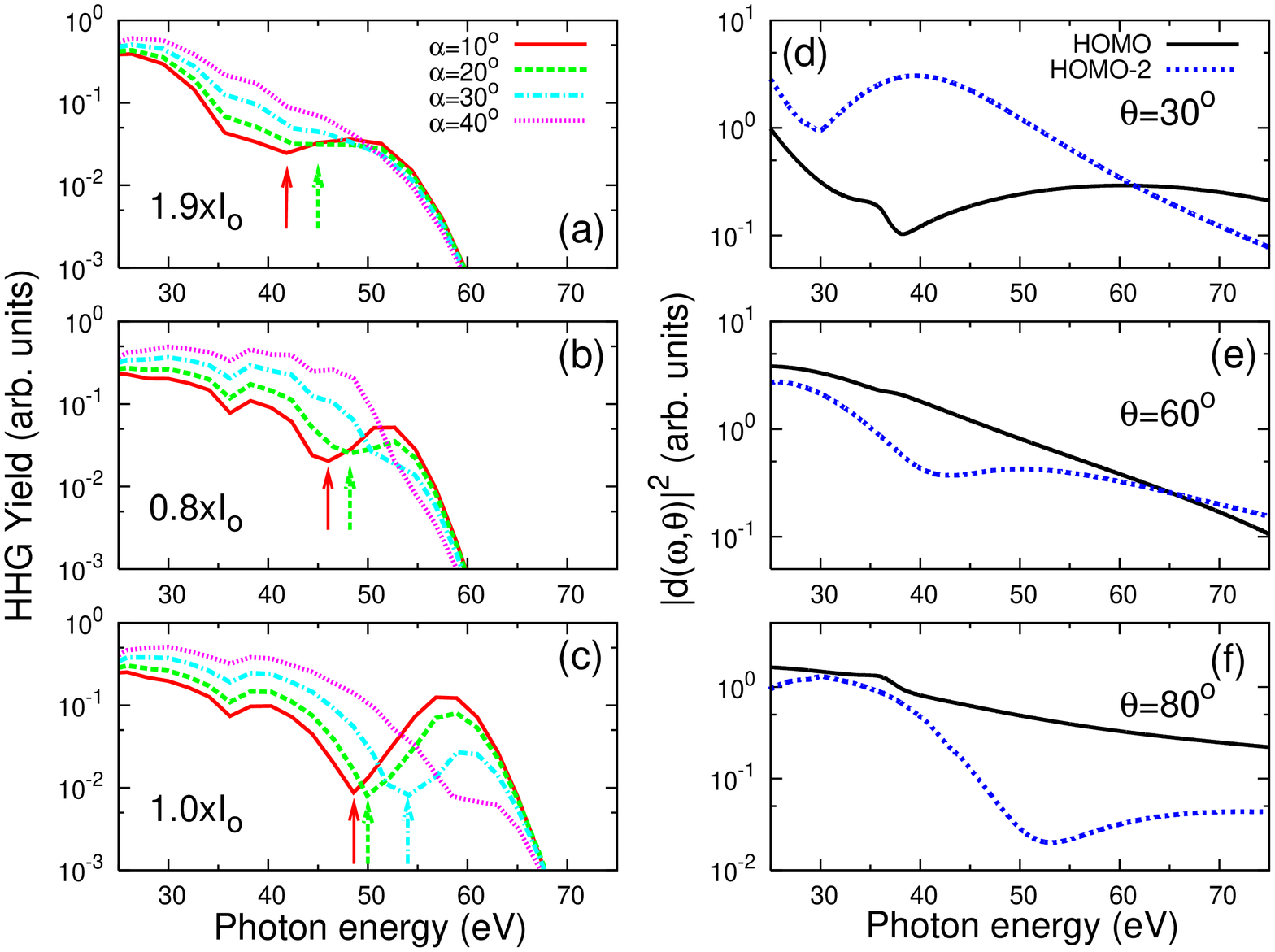}}}}
\caption{(Color online) Dependence of macroscopic HHG spectra
(envelope only) on pump-probe angles $\alpha$=10$^{\circ}$,
20$^{\circ}$, 30$^{\circ}$, and 40$^{\circ}$ (a) for an 800-nm
laser, (b) and (c) for a 1200-nm laser. Laser intensities are given
in units of I$_{\circ}$=10$^{14}$ W/cm$^{2}$. Arrows indicate the
positions of minima. Degree of alignment:
$\langle\cos^{2}\theta\rangle$=0.60. (d)-(f) Photorecombination
transition dipoles (parallel component, the square of magnitude) of
HOMO and HOMO-2 in terms of photon energy at alignment angles
$\theta$=30$^{\circ}$, 60$^{\circ}$, and 80$^{\circ}$. \label{Fig7}}
\end{figure*}

In W\"{o}rner {\it et al.} \cite{hans-prl-10}, it was found that the
minimum in the HHG spectra of aligned CO$_{2}$ moved to higher
photon energy with increasing pump-probe angle $\alpha$, i.e., the
angle between aligning pump beam and the HHG generating probe beam
polarizations. This phenomenon has also been observed in other
measurements \cite{Mairesse-jmo-2008,olga-nature-2009}. In contrast,
the minimum in the HHG spectra of aligned N$_{2}$ has been found to
stay at the same position for all pump-probe angles
\cite{Mairesse-jmo-2008,Haessler-NatPhys-2010}.

We show the calculated HHG spectra at four pump-probe angles in an
800-nm laser in Fig.~{\ref{Fig7}}(a), and in a 1200-nm laser in
Figs.~{\ref{Fig7}}(b) and {\ref{Fig7}}(c). Laser parameters and
experimental arrangements are the same as in Fig.~{\ref{Fig1}} and
the degree of alignment is $\langle\cos^{2}\theta\rangle$=0.60. The
HHG spectra for $\alpha$=0$^{\circ}$ have been shown in
Fig.~{\ref{Fig3}}. These figures show that the minimum in the HHG
spectra moves to higher photon energies with increasing $\alpha$ ,
and the minimum disappears at large angles. At larger pump-probe
angles, contributions from large alignment angles increase. Since
HOMO dominates over HOMO-2 at large angles in both the ionization
rates [see Figs.~{\ref{Fig5}}(c) and {\ref{Fig5}}(d)] and the PR
transition dipoles [see Figs.~{\ref{Fig7}}(d)-{\ref{Fig7}}(f)], thus
HHG at large pump-probe angles has essentially no contributions from
HOMO-2. Also note that at small $\alpha$, the total HHG yield is
much smaller \cite{at-pra-2009}. In fact, the total HHG spectra for
randomly distributed CO$_2$ have little contributions from molecules
that are aligned nearly parallel to the polarization axis of the
laser. The HHG spectra of CO$_2$ are complex only in the region
where HHG yields are small. This is generally true -- interpretation
of small processes always requires careful and detailed theories.

\section{Summary and outlook}

In this paper we have analyzed the multiple orbital contribution to
HHG in CO$_2$ with the inclusion of macroscopic propagation effect.
In the past few years, there have been many experimental and
theoretical studies on the HHG of CO$_2$ from many laboratories,
using lasers with different wavelengths and intensities, for CO$_2$
molecules that are randomly distributed or partially aligned. In
particular, for CO$_2$ molecules that are partially aligned along
the polarization axis of the probe laser, many experiments have
shown that the HHG spectra exhibit minima and the positions of the
minima shift  with laser intensities
\cite{olga-nature-2009,hans-prl-10,Torres-pra-2010}. The shift of
the minimum position with laser intensities has been attributed to
the interference between the contributions to HHG from the HOMO-2
with the one from the HOMO, despite HOMO-2 lying at 4.4 eV deeper
than the HOMO. Since HHG is a nonlinear process, these observations
posed a great challenge to the theory, especially for the prediction
of the position of the minimum and how it changes with the
experimental conditions. Since all experimental HHG spectra include
macroscopic propagation effect, comparison of theory with experiment
is incomplete unless the theory also has included the propagation
effect. Our analysis in this paper is based on the macroscopic
propagation code extended for aligned molecules, and the recently
developed quantitative rescattering theory. We find that although
HHG spectra change significantly under different experimental
parameters such as degree of alignment, focussing condition, and the
use of a slit, the position of the minimum in the HHG spectra
behaves in a similar trend as laser intensity and pump-probe angle
vary. This trend has been found to be consistent with the recent
experimental measurements from different groups.

We comment that the present theory and the earlier one by Smirnova
{\it et al.} \cite{olga-nature-2009} both explain the intensity
dependence of the change of HHG minima, but the details between the
two theories are quite different. The alignment dependence of the
ionization rates, the recombination dipole matrix elements and their
phases entering the two theories are not the same, for both the HOMO
and HOMO-2. As illustrated in this paper, these parameters can all
affect the position of the predicted interference minimum.
Furthermore, in Smirnova {\it et al.} \cite{olga-nature-2009} the
interference is attributed to the importance of hole dynamics in the
ion core. Our approach is formulated in the time-independent
fashion, no hole dynamics is included. Since HHG spectra are taken
without explicit observation of electron dynamics, the difference
between the two models cannot be settled. Despite these differences,
our understanding of the HHG spectra of CO$_2$ has come a long way
since 2005 \cite{Kanai-nature-2005}. With the possibility of
including macroscopic propagation effect ``routinely" in the HHG
theory for molecular targets, further experimental studies should
explore the effects of laser focusing condition and gas pressure,
for lasers extending to longer wavelengths. Such studies would
further our basic understanding of strong-field physics of molecules
to the next level, and eliminate the need of introducing extraneous
assumptions in the interpretation of experimental data.

\section{Acknowledgments}
This work was supported in part by Chemical Sciences, Geosciences
and Biosciences Division, Office of Basic Energy Sciences, Office of
Science, U.S. Department of Energy.

\end{document}